\documentclass[showpacs,twocolumn,amsmath,amssymb,superscriptaddress,preprintnumbers,nofootinbib,showkeys]{revtex4}

\begin{document}

\newcommand{\nablab}{{\mathop {\rule{0pt}{0pt}{\nabla}}\limits^{\bot}}\rule{0pt}{0pt}}

\title{The extended Einstein-Maxwell-aether-axion theory: \\ Effective metric as an instrument of the aetheric control over the axion dynamics}

\author{Alexander B. Balakin}
\email{Alexander.Balakin@kpfu.ru} \affiliation{Department of
General Relativity and Gravitation, Institute of Physics, Kazan
Federal University, Kremlevskaya str. 16a, Kazan 420008, Russia}

\author{Amir F. Shakirzyanov}
\email{shamirf@mail.ru} \affiliation{Department of
General Relativity and Gravitation, Institute of Physics, Kazan
Federal University, Kremlevskaya str. 16a, Kazan 420008, Russia}

\date{\today}

\begin{abstract}
In the framework of the Einstein-Maxwell-aether-axion theory we consider the self-consistent model based on the concept of a two-level control, which is carried out by the dynamic aether over the behavior of the axionically active electrodynamic system. The Lagrangian of this model contains two guiding functions, which depend on four differential invariants of the aether velocity: the scalar of expansion of the aether flow, the square of the acceleration four-vector, the squares of the shear and vorticity tensors. The guiding function of the first type is an element of the effective aetheric metric; this effective metric is involved in the formulation of kinetic terms for the vector, pseudoscalar and electromagnetic fields and predetermines features of their evolution. The guiding function of the second type is associated with the distribution of axions and describes its vacuum average value; basically, this function appears in the potential of the axion field and predetermines the position and depth of its minima.  The self-consistent set of coupled master equations of the model is derived. The example of the static spherically symmetric system is considered as an application.

\end{abstract}
\maketitle

\section{Introduction}

The term {\it effective metric} includes information about several known geometric objects. The first representative of the class of effective metrics is the so-called Gordon's optical metric, $g_{*}^{ik}=g^{ik} {+} (n^2{-}1) V^i V^k$, which was introduced a hundred years ago in 1923 \cite{Gordon} for arbitrary moving isotropic, transparent, nondispersive dielectric medium with the constant refraction index $n$. The four-vector $V^k$ describes the velocity of this medium, and $g^{ik}$ relates to the physical spacetime metric. Due to the interaction with the medium, photons do not follow the null geodesic lines in the physical spacetime with the metric $g_{ij}$, however, the worldlines of these photons coincide with  the null geodesic lines in the fictitious spacetime with the optical metric $g_{*ij}$. These results, initially fair within the framework of the approximation of geometric optics \cite{GR}, were generalized based on the formalism of covariant phenomenological electrodynamics of continuous media \cite{PMQ,MauginJMP}. It turns out that the Gordon's optical metric is related to the so-called {\it associated} metric, which allows us to rewrite the Tamm's constitutive tensor \cite{HehlObukhov,LL} in a quasi-vacuum form (see, e.g., \cite{BZ} for details).
The second representative of the class of effective metrics is the {\it acoustic} metric \cite{Visser1,Novello,VolovikBook}. Using this acoustic metric, one can
regard sound waves as quasi-particles moving in the effective spacetime (see, e.g., \cite{Volovik}). The third example of the effective metric was described in \cite{BDZ1,BDZ2,BDZ3}; these works dealt with the {\it color} metrics, {\it color-acoustic} and associated metrics, which appear in the SU(N) symmetric Einstein-Yang-Mills-Higgs theory.

New aspects of the effective metric formalism can be opened, if we work with the theory of dynamic aether \cite{J1,J2,J3,J4,J5}, which is characterized by the unit timelike vector field $U^j$ associated with the velocity four-vector of the aether. Formally speaking, the macroscopic velocity of the dielectric medium, $V^m$, is defined algebraically, e.g., as the timelike eigenvector of the stress-energy tensor, while the four-vector $U^j$ is defined dynamically, i.e., on the language of the field theory. Nevertheless, one can define the effective aetheric metric $G^{ik}= g^{ik} {+} {\cal H} U^i U^k$ in analogy with $g_{*}^{ik}$. In this context one new scalar ${\cal H}$ appears in the theory, thus extending the set of instruments for modeling.

How do we plan to use this aetheric metric in the Einstein-Maxwell-aether-axion theory extension? This theory contains three so-called constitutive tensors, based on which the kinetic terms for the vector, axion and electromagnetic fields are formulated. The first one is the Jacobson's tensor $K^{ab}_{\ \ mn}$, appeared in the Lagrangian in the convolution $K^{ab}_{\ \ mn} \nabla_a U^m \nabla_b U^n$ ($\nabla_k$ denotes the covariant derivative). The second constitutive tensor ${\cal G}^{mn}$ appears in the kinetic terms of the axion field  ${\cal G}^{mn}\nabla_m \phi \nabla_n \phi$. The constitutive Tamm's tensor $C^{ikmn}$ is the standard element of the scalar $C^{ikmn}F_{ik} F_{mn}$, where $F_{mn}$ is the Maxwell tensor. We suggest to modify these three constitutive tensors by replacing the physical metric $g^{pq}$, which enters these tensors, by the aetheric effective metric $G^{mn}= g^{mn} + {\cal H} U^m U^n$. We have to emphasize that the Christoffel symbols, Riemann and Ricci tensors are still based on the spacetime metric $g_{pq}$.
Below we indicate the function ${\cal H}$ as the guiding function of the first type. Why do we use the term guiding function in this context? First, we assume that ${\cal H}$ depends on time and coordinates  via the scalars, which are constructed using the covariant derivative of the aether velocity four-vector, $\nabla_k U^j$, only. Second, we will see below that the function ${\cal H}$ predetermines the evolution (and/or distribution) of the axion field, as well as, the properties of the permittivity tensors of the aetherically active electrodynamic system.

When we describe the potential of the axion field, we introduce the aetheric guiding function of the second type, $\Phi_*$, which is associated with the vacuum average distribution of the axions. This function also depends on the differential invariants of the aether velocity and describes the position and depth of the minima of the axion potential.

The paper is organized as follows. In Section II we develop the mathematical formalism of the extended Einstein-Maxwell-aether-axion theory, and derive the set of coupled master equations for the vector, pseudoscalar, electromagnetic and gravitational fields. In Section III we consider the application of the elaborated formalism to the model describing the static spherically symmetric axionic dyon. Section IV  contains discussion and conclusions.

\section{The formalism}

\subsection{The action functional}

We consider the model of interaction of four fields: the gravitational and unit vector fields are the key elements of the Einstein-aether theory \cite{J1,J2,J3,J4}; the pseudoscalar field appears to describe the axionic dark matter \cite{PQ} - \cite{a5}; the electromagnetic field is considered to be coupled to the aether, to the axion and gravitational fields. Also, we are ready to consider a medium, which interacts with the aether and can be associated, for instance, with non-axionic dark matter, cosmic dark energy, dust, baryon matter, etc. The total action functional is presented by the following sum:
\begin{equation}
{-}S_{(\rm total)} {=} \int d^4x \sqrt{{-}g} \left\{ L_{(\rm EA)} {+} L_{(\rm A)} {+} L_{(\rm EM)} {+} L_{(\rm M)} \right\},
\label{0}
\end{equation}
where $g$ is the determinant of the spacetime metric $g_{mn}$.

\subsubsection{Extended Lagrangian of the Einstein-aether theory}

We work with the standard Lagrangian
$$
L_{(\rm EA)} = \frac{1}{2\kappa}\left[R{+}2\Lambda {+} \lambda\left(g_{mn}U^mU^n {-} 1 \right) {+}  \right.
$$
\begin{equation}
\left.
+ K^{ab}_{\ \ mn} \nabla_a U^m  \nabla_b U^n \right] \,,
\label{01}
\end{equation}
which contains the Ricci scalar $R$, the cosmological constant $\Lambda$, the Einstein constant  $\kappa$, the Lagrange multiplier $\lambda$, the unit timelike vector field $U^i$, which is associated with the velocity four-vector of the aether flow, and the covariant derivative $\nabla_k$, in which the connection $\Gamma^j_{mn}$ is consistent with the spacetime metric $g_{mn}$, i.e., $\nabla_k g_{mn}=0$.
The constitutive tensor proposed by Jacobson and colleagues
$$
K^{ab}_{\ \ mn} = C_1 g^{ab}g_{mn} {+} C_2 \delta^{a}_{m} \delta^{b}_{n} {+}
$$
\begin{equation}
+C_3 \delta^{a}_{n} \delta^{b}_{m} {+} C_4 U^a U^b g_{mn}
\label{11}
\end{equation}
contains four phenomenological constants $C_1$, $C_2$, $C_3$, $C_4$. If we use the transformation $g^{mn} \to G^{mn}$, where $G^{mn}= g^{mn} {+} {\cal H}U^m U^n$ is the effective metric, the scalar ${\cal K} \equiv K^{ab}_{\ \ mn}(\nabla_a U^m) (\nabla_b U^n)$ keeps the form, however, the constant $C_4$ has to be replaced by $\tilde{C}_4 \equiv C_4+{\cal H} C_1$ (we keep in mind that $U^q \nabla_m U_q = 0$ due to the normalization condition $g_{mn}U^m U^n =1$). In other words, when the term ${\cal H}$ is constant, the transformation $g^{mn} \to G^{mn}$ does not change the structure of the
scalar ${\cal K}$, and we deal with the redefinition of the phenomenological constant $C_4$.

Now we consider ${\cal H}$ to be a function.
What can be organized the structure of the scalar ${\cal H}$? We assume that this quantity depends on four scalars constructed using the covariant derivative of the aether velocity four-vector ${\cal H}(\Theta, a^2, \sigma^2, \omega^2)$, and we have to discuss in more details these four arguments of the guiding function of the first type.
Using the unit four-vector $U^j$ one can decompose all the tensor quantities into the sum of the so-called longitudinal and transversal components. In particular, the covariant derivative can be decomposed as follows:
\begin{equation}
\nabla_k = U_k D {+} \nablab_k \,, \quad D = U^s \nabla_s \,, \quad \nablab_k = \Delta_k^j \nabla_j \,,
\label{F134}
\end{equation}
$\Delta_k^j {=} \delta^j_k {-} U^j U_k $ is the projector. The covariant derivative $\nabla_k U_j$ can be decomposed as
\begin{equation}
\nabla_k U_j = U_k DU_j + \sigma_{kj} + \omega_{kj} + \frac13 \Delta_{kj} \Theta \,,
\label{F54}
\end{equation}
where the acceleration four-vector $DU_j$, the symmetric traceless shear tensor $\sigma_{kj}$, the skew - symmetric vorticity tensor $\omega_{kj}$ and the expansion scalar $\Theta$ are presented by the well-known formulas
$$
DU_j = U^s \nabla_s U_j \,, \quad \sigma_{kj} = \frac12 \left(\nablab_k U_j {+} \nablab_j U_k \right) {-} \frac13 \Delta_{kj} \Theta \,,
$$
\begin{equation}
\omega_{kj} = \frac12 \left(\nablab_k U_j {-} \nablab_j U_k \right) \,, \quad
\Theta = \nabla_kU^k \,.
\label{F541}
\end{equation}
The decomposition (\ref{F54}) allows us to introduce one linear and three quadratic scalars
$$
\Theta = \nabla_k U^k \,, \quad a^2 = DU_k DU^k \,,
$$
\begin{equation}
\sigma^2 = \sigma_{mn} \sigma^{mn} \,, \quad \omega^2 = \omega_{mn} \omega^{mn} \,,
\label{06}
\end{equation}
and to consider them as the arguments of the scalar function ${\cal H}$. In these terms the scalar ${\cal K}$ can be rewritten in more convenient form
$$
{\cal K} \equiv K^{ab}_{\ \ mn}(\nabla_a U^m) (\nabla_b U^n) =
 [C_1(1+{\cal H}) {+} C_4] a^2 {+}
 $$
 \begin{equation}
{+}
(C_1 {+} C_3)\sigma^2 {+}
(C_1 {-} C_3)\omega^2 {+} \frac13 \left(C_1 {+} 3C_2 {+}C_3 \right) \Theta^2
\,. \label{act5n}
\end{equation}
Taking into account the constraints obtained after the detection of the event GRB170817A \cite{GW}, we have to put $C_1{+}C_3=0$; below we work with the modified Jacobson's invariant
\begin{equation}
{\cal K} =
 [C_1(1+{\cal H}) {+} C_4] a^2 {+} 2C_1 \omega^2 {+} C_2 \Theta^2 \,, \label{2act5n}
\end{equation}
in which the shear tensor $\sigma_{mn}$ happens to be hidden.

\subsubsection{Extended Lagrangian of the axion field}

We use here the following generalization: we take the standard Lagrangian of the pseudoscalar field
\begin{equation}
\frac12 \Psi^2_0   \left[V(\phi) {-} g^{mn}\nabla_m \phi \nabla_n \phi \right]
\label{131}
\end{equation}
and replace, first, the spacetime metric $g^{mn}$ with the effective metric $G^{mn}$; second, the axion field potential $V(\phi)$ with the modified potential $V(\phi, \Phi_*)$, assuming that the  guiding function of the second type, $\Phi_*$, depends on four arguments introduced above, i.e., $\Phi_*(\Theta, a^2, \sigma^2, \omega^2)$. We obtain now the modified Lagrangian of the axion field
\begin{equation}
L_{(\rm A)} = \frac12 \Psi^2_0   \left[V(\phi, \Phi_*) {-} \left(g^{mn} + {\cal H} U^m U^n \right)\nabla_m \phi \nabla_n \phi \right] \,.
\label{132}
\end{equation}
We choose the potential of the axion field in the periodic form
\begin{equation}
V(\phi,\Phi_{*}) = \frac{m^2_A \Phi^2_{*}}{2\pi^2} \left[1- \cos{\left(\frac{2 \pi \phi}{\Phi_{*}}\right)} \right] \,,
\label{13}
\end{equation}
thus inheriting the discrete symmetry $\frac{2\pi \phi}{\Phi_*} \to \frac{2\pi \phi}{\Phi_*} + 2\pi k$. The parameter $\Psi_0$ relates to the coupling constant of the
axion-photon interaction $g_{A \gamma \gamma}$ ($\frac{1}{\Psi_0}=g_{A \gamma \gamma}$).
This periodic potential has the minima at $\phi=n \Phi_{*}$. Near the minima, when
$\phi \to n \Phi_{*} {+} \psi$ and $|\frac{2\pi \psi}{\Phi_*}|$ is small, the potential takes the standard form $V \to m^2_A \psi^2$, where $m_A$ is the axion rest mass.
As it was advocated in \cite{Equilib1,Equilib2,Equilib3,Equilib4,Equilib5}), we deal with the axionic analog of the equilibrium state, when $\phi {=} n \Phi_*$, since
$V_{|\phi{=}n\Phi_*} {=}0$, and $\left(\frac{\partial V}{\partial \phi} \right)_{|\phi{=}n\Phi_*} {=}0$. Should be mentioned that there exist an alternative terminology for this state, namely, frozen axions (see, e.g., \cite{oiko1,oiko2}). This terminology is advocated by the fact that in terms of equations of state such axion configuration can be described by the equation $P=W$ with formally defined velocity of sound coinciding with the speed of light.

\subsubsection{Extended Lagrangian of the electromagnetic field}

We generalize the standard Lagrangian of the electromagnetic field coupled to the pseudoscalar field \cite{WTNi77}
\begin{equation}
L_{(\rm EM)} =   \frac18 \left[g^{mp} g^{nq} - g^{mq} g^{np} +  \phi \epsilon^{mnpq} \right] F_{mn}F_{pq} \,,
\label{71}
\end{equation}
using replacements of two types: first,  $g^{mn} \to G^{mn}$, second,  $\phi \to \frac{\Phi_*}{2\pi}\sin{\left(\frac{2\pi \phi}{\Phi_*}\right)}$. The first replacement was discussed above, and we have to focus now on the second one. First, when $\phi \to 0$, we obtain the standard $\phi$  from the sinusoidal term. Second, the sinus is the odd function, as it is necessary for the description of the pseudoscalar field. Third, sinus is periodic function, and we keep the discrete symmetry prescribed for the axion field.

Keeping in mind these arguments, we obtain the modified Lagrangian of the axion field
$$
{\cal L}_{(\rm EM)} =   \frac14 F_{mn} F^{mn} {+} \frac12 {\cal H} F_{mn} U^n F^{m}_{ \ \ q}U^q  {+}
$$
\begin{equation}
+ \frac{\Phi_*}{8\pi}\sin{\left(\frac{2\pi \phi}{\Phi_*}\right)} F^{*mn}F_{mn}\,,
\label{72}
\end{equation}
where $F_{mn}$ is the Maxwell tensor, $F^{*mn} \equiv \frac12 \epsilon^{mnpq} F_{pq}$ is its dual and $\epsilon^{mnpq}=\frac{E^{mnpq}}{\sqrt{-g}}$ is the Levi-Civita tensor
(we use the definition $E^{0123}=1$ for the Levi-Civita symbol $E^{mnpq}$).
This part of the Lagrangian can be rewritten in terms of the Tamm tensor $C^{mnpq}$ as follows:
\begin{equation}
{\cal L}_{(\rm EM)} =  \frac14 C^{mnpq}F_{mn} F_{pq} \,,
\label{73}
\end{equation}
where the Tamm tensor
$$
C^{mnpq} = \frac12 \left[\left(g^{mp} g^{nq}{-}g^{mq}g^{np} \right) {+}  \frac{\Phi_*}{2\pi}\sin{\left(\frac{2\pi \phi}{\Phi_*}\right)} \epsilon^{mnpq} {+}
\right.
$$
\begin{equation}
\left. {+} {\cal H}\left(g^{mp}U^n U^q {-} g^{mq}U^n U^p {+} g^{nq}U^m U^p {-} g^{np}U^m U^q  \right) \right]
\label{74}
\end{equation}
contains both aetheric guiding functions: ${\cal H}$ and $\Phi_*$.

Mention should be made that in the electrodynamics of isotropic homogeneously moving continua \cite{MauginJMP,HehlObukhov} the coefficient ${\cal H}$ has a direct interpretation, ${\cal H}= n^2{-}1$, where $n = \sqrt{\varepsilon \mu}$ is the refraction index of the medium. In this sense the function $\sqrt{1{+} {\cal H}}$ plays the role of an aetheric refraction index.

Finally,
we assume that the Lagrangian of the medium $L_{(\rm M)}$ does not depend on the aether velocity $U^j$, on the axion field $\phi$, on the Maxwell tensor $F_{mn}$
and its dual $F^*_{mn}$, however, it can depend on the potential of the electromagnetic field $A_k$ ($F_{mn}= \nabla_m A_n {-}\nabla_n A_m$), if the medium possesses free electric charges and produces the electric current.

Keeping in mind all discussed modifications, we obtain the following extended total action functional
$$
-S_{(\rm total)} = \int d^4x \sqrt{-g} \left\{ \frac{1}{2\kappa}\left[R{+}2\Lambda {+} \lambda\left(g_{mn}U^mU^n {-} 1 \right) {+} \right. \right.
$$
$$
\left. \left. {+}\left(C_1(1{+}{\cal H}) {+} C_4 \right) a^2 {+} 2C_1 \omega^2 {+} C_2 \Theta^2 \right] {+}
\right.
$$
$$
\left. + \frac12 \Psi^2_0   \left[\frac{m^2_A \Phi^2_{*}}{2\pi^2} \left[1- \cos{\left(\frac{2 \pi \phi}{\Phi_{*}}\right)} \right] {-} \nabla_m \phi \nabla^m \phi {-} {\cal H} (D \phi)^2 \right] +
\right.
$$
$$
\left. + \frac14 F_{mn} F^{mn} + \frac12 {\cal H} F_{mn} U^n F^{m}_{ \ \ q}U^q  +  \right.
$$
\begin{equation}
\left. + \frac{\Phi_*}{8\pi}\sin{\left(\frac{2\pi \phi}{\Phi_*}\right)} F^{*mn}F_{mn}
+  L_{(\rm M)} \right\} \,.
\label{00}
\end{equation}
The standard variation procedure gives us the master equations of the model.

\subsection{Master equations of the model}

\subsubsection{Master equations for the unit vector field}

Variations of the total action functional (\ref{00}) with respect to the Lagrange multiplier $\lambda$ gives the normalization condition
\begin{equation}
g_{mn}U^m U^n =1 \,.
\label{14}
\end{equation}
Variation with respect to the four-vector $U^i$ gives the following set of equations:
\begin{equation}
\nabla_a {\cal J}^{a}_{\ j}  = \lambda  U_j  {+}  \left(C_4 + C_1 {\cal H} \right) DU_m \nabla_j U^m {+}
\label{15}
\end{equation}
$$
+ {\cal H}\left[-\kappa \Psi^2_0  D \phi \nabla_j \phi {+} \kappa  F_{mj} F^{mq} U_q \right]-
$$
$$
{-} \nabla_j \left(\Omega_1 \frac{\partial \Phi_*}{\partial \Theta} {+} \Omega_2 \frac{\partial {\cal H}}{\partial \Theta} \right) -
$$
$$
{-}2DU_j D\left(\Omega_1 \frac{\partial \Phi_*}{\partial a^2} {+} \Omega_2 \frac{\partial {\cal H}}{\partial a^2} \right) {-}
$$
$$
-2\nabla^n \left[\left(\Omega_1 \frac{\partial \Phi_*}{\partial \sigma^2}  {+} \Omega_2 \frac{\partial {\cal H}}{\partial \sigma^2} \right)\sigma_{jn}\right]
{+}
$$
$$
{+}2\nabla^n \left[\left(\Omega_1 \frac{\partial \Phi_*}{\partial \omega^2}  {+} \Omega_2 \frac{\partial {\cal H}}{\partial \omega^2}\right)\omega_{jn}\right] +
$$
$$
 {+} 2\left(\Omega_1 \frac{\partial \Phi_*}{\partial a^2} {+} \Omega_2 \frac{\partial {\cal H}}{\partial a^2}\right)  \left(DU_k \nabla_j U^k {-} \Theta DU_j {-}D^2 U_j \right) {-}
$$
$$
-2\left(\Omega_1 \frac{\partial \Phi_*}{\partial \sigma^2} {+} \Omega_2 \frac{\partial {\cal H}}{\partial \sigma^2} \right) DU^n \sigma_{jn} -
$$
$$
-2\left(\Omega_1 \frac{\partial \Phi_*}{\partial \omega^2}{+} \Omega_2 \frac{\partial {\cal H}}{\partial \omega^2}\right) DU^n \omega_{jn} \,.
$$
Here we introduced the following auxiliary definitions:
$$
\Omega_1 {=} \frac{\kappa \Psi_0^2 m_A^2}{2\pi^2} \left\{\Phi_* \left[1{-} \cos{\left(\frac{2 \pi \phi}{\Phi_{*}}\right)}\right] {-}\pi \phi \sin{\left(\frac{2 \pi \phi}{\Phi_{*}}\right)} \right\} {+}
$$
\begin{equation}
 {+}
\frac{\kappa}{8\pi} F^*_{mn}F^{mn} \left[\sin{\left(\frac{2 \pi \phi}{\Phi_{*}}\right)} {-} \frac{2\pi \phi}{\Phi_*} \cos{\left(\frac{2 \pi \phi}{\Phi_{*}}\right)} \right] \,,
\label{Omega1}
\end{equation}
\begin{equation}
\Omega_2 {=} \frac12 C_1 DU_m DU^m {-} \frac12 \kappa \Psi^2_0 (D \phi)^2 {+} \frac12 \kappa F_{mn} U^n F^{mq}U_q   \,.
\label{Omega2}
\end{equation}
The tensor ${\cal J}^{a}_{ \ j}$ is now of the form
$$
{\cal J}^{a}_{ \ j} = K^{ab}_{\ \ jn} \nabla_b U^n
= C_1 \left( \nabla^a U_j - \nabla_j U^a \right) +
$$
\begin{equation}
+ C_2 \delta^a_j \Theta  + (C_4 + C_1 {\cal H}) U^a DU_j  \,.
\label{16}
\end{equation}
Convolution of (\ref{15}) with $U^j$ gives us the function $\lambda$:
$$
\lambda =  U^j \nabla_a {\cal J}^{a}_{\ j}  {-} \left( C_4 + {\cal H} C_1 \right)DU_m DU^m  {+}
$$
\begin{equation}
{+}{\cal H}\left[ \kappa \Psi^2_0  (D \phi)^2 {-} \kappa  F_{mj}U^j F^{mq} U_q \right]+
\label{150}
\end{equation}
$$
{+} D \left(\Omega_1 \frac{\partial \Phi_*}{\partial \Theta} {+} \Omega_2 \frac{\partial {\cal H}}{\partial \Theta} \right)
 {-} 2 \sigma^2 \left(\Omega_1 \frac{\partial \Phi_*}{\partial \sigma^2}  {+} \Omega_2 \frac{\partial {\cal H}}{\partial \sigma^2} \right){-}
$$
$$
{-}2 \omega^2 \left(\Omega_1 \frac{\partial \Phi_*}{\partial \omega^2}  {+} \Omega_2 \frac{\partial {\cal H}}{\partial \omega^2}\right) -
 4 a^2 \left(\Omega_1 \frac{\partial \Phi_*}{\partial a^2} {+} \Omega_2 \frac{\partial {\cal H}}{\partial a^2}\right)  \,.
$$

\subsubsection{Master equation for the axion field}

Variation of the total action functional with respect to the axion field yields
$$
\nabla_m \left[\left(g^{mn} + {\cal H} U^m U^n \right)\nabla_n \phi \right] + \frac{m^2_A \Phi_{*}}{2\pi} \sin{\left(\frac{2 \pi \phi}{\Phi_{*}}\right)}  =
$$
\begin{equation}
= - \frac{1}{4\Psi_0^2} \cos{\left(\frac{2\pi \phi}{\Phi_*}\right)} F^*_{mn}F^{mn} \,,
\label{24}
\end{equation}
or in more detail
$$
 (1{+} {\cal H}) D^2 \phi {+} [(1{+}{\cal H})\Theta + D {\cal H}]D\phi {-} DU^m  \nablab_m \phi {+} \nablab_m \nablab^m \phi {+}
$$
\begin{equation}
{+} \frac{m^2_A \Phi_{*}}{2\pi} \sin{\left(\frac{2 \pi \phi}{\Phi_{*}}\right)}  {=} {-} \frac{1}{4\Psi_0^2} \cos{\left(\frac{2\pi \phi}{\Phi_*}\right)} F^*_{mn}F^{mn} \,.
\label{243}
\end{equation}

\subsubsection{Master equations for the electromagnetic field}

The first subset of the electrodynamic equations is standard
\begin{equation}
\nabla_k F^{*ik} =0 \,.
\label{T0}
\end{equation}
Variation of the action functional (\ref{00}) with respect to the electromagnetic potential gives the second set of electrodynamic equations
\begin{equation}
\nabla_n \left[C^{mnpq} F_{pq} \right]=0 \,,
\label{T1}
\end{equation}
where the Tamm constitutive tensor $C^{mnpq}$, which depends on the metric $g^{mn}$, on the aether velocity four-vector $U^k$, on the guiding scalars $\Theta$, $a^2$, $\sigma^2$, $\omega^2$, and on the axion field $\phi$,  is presented by (\ref{74}). When one works with the electric field four-vector $E^p =F^{pq} U_q$ and with the magnetic excitation four-vector $B_p = F^{*}_{pq} U^q$, one assumes that the aetheric velocity is an analog of the medium velocity in the linear electrodynamics of continua \cite{HehlObukhov}. It is well known that these four-vectors, orthogonal to the velocity four-vector $U^j$, appear in the decompositions
$$
F^{mn} = E^m U^n - E^n U^m - \epsilon^{mnpq} B_p U_q\,,
$$
\begin{equation}
F^{* mn} = B^m U^n - B^n U^m + \epsilon^{mnpq} E_p U_q \,.
\label{T2}
\end{equation}
Also, we know that the Tamm tensor predetermines the dielectric permittivity tensor $\varepsilon^{mp}$, the magnetic impermeability tensor $(\mu^{-1})_{pq}$ and the tensor magneto-electric coefficients $\nu^{mp}$. For the presented model we obtain
$$
\varepsilon^{mp} = 2 C^{mnpq}U_n U_q = (1+ {\cal H})\Delta^{mp} \,,
$$
$$
(\mu^{-1})_{pq} = - \frac12 \epsilon_{psik} C^{ikmn} \epsilon_{mnql} U^s U^l = \Delta_{pq} \,,
$$
\begin{equation}
\nu_i^{\ p} = \epsilon_{ismn} C^{mnpq} U^s U_q = -  \frac{\Phi_*}{2\pi} \Delta^p_i \ \sin{\left(\frac{2\pi \phi}{\Phi_*}\right)}\,.
\label{T10}
\end{equation}
In other words, the presented electrodynamic system coupled to the aether and axion field behaves as a medium with the magnetic permeability $\mu=1$, dielectric permittivity $\varepsilon = 1 {+} {\cal H}$, and symmetric tensor of the magnetoelectric coefficient (\ref{T10}). The novelty of the last formula is that the magnetoelectricity disappears if the axion system is in the equilibrium state, i.e., $\phi= n \Phi_*$.

\subsubsection{Master equations for the gravitational field}

Variation of the action functional (\ref{00}) with respect to the metric gives the gravity field equations:
$$
	R_{ik} - \frac12 R g_{ik} - \Lambda g_{ik} =
$$
\begin{equation}
= T^{(\rm U)}_{ik} + \kappa T^{(\rm A)}_{ik} + \kappa T^{(\rm EM)}_{ik} +  T^{(\rm INT)}_{ik} + \kappa T^{(\rm M)}_{ik}\,.
	\label{25}
\end{equation}
We indicate the first term in the right-hand side of (\ref{25}) as the stress-energy tensor associated with the aether flow; it contains the following elements:
$$
	T^{(\rm U)}_{ik} =
	\frac12 g_{ik} \ K^{abmn} \nabla_a U_m \nabla_b U_n {+}
$$
$$
{+}\nabla^m \left[U_{(i}{\cal J}_{k)m} {-}
{\cal J}_{m(i}U_{k)} {-}
{\cal J}_{(ik)} U_m\right]{+} U_iU_k U_j \nabla_a {\cal J}^{aj} {+}
$$
$$
{+}C_1\left[(\nabla_mU_i)(\nabla^m U_k) {-}
(\nabla_i U_m )(\nabla_k U^m) \right] {+}
$$
\begin{equation}
{+}(C_4 {+} C_1 {\cal H})\left( D U_i D U_k {-} U_iU_k DU_m DU^m \right)\,.
\label{326}
\end{equation}
The parentheses symbolize the symmetrization of indices.
The second term is associated with the stress-energy tensor of the axion field; we include the following elements to this  construction:
$$
	T^{(\rm A)}_{ik}= \Psi^2_0 \left[\nabla_i \phi \nabla_k \phi  {+} {\cal H} (D \phi)^2 \left(U_iU_k {-}\frac12 g_{ik}\right) + \right.
$$
\begin{equation}
\left. + \frac12 g_{ik}\left(V {-} \nabla_s \phi \nabla^s \phi  \right) \right]\,.
		\label{T76}
\end{equation}
The third term is attributed to the stress-energy tensor of the electromagnetic field:
$$
 T^{(\rm EM)}_{ik}=  \left( \frac14 g_{ik} F_{mn} F^{mn} - F_{in} F_{k}^{\ n} \right) +
 $$
 \begin{equation}
 + {\cal H}\left[\left(\frac12 g_{ik}- U_i U_k \right)E_m E^m - E_i E_k  \right]\,.
\label{T91}
\end{equation}
This tensor is symmetric and traceless.
Other elements of the right-hand side of the equations of the gravity field are regrouped into the so-called interaction term; it contains all the derivatives of the guiding functions $\Phi_*$ and ${\cal H}$ with respect to their arguments $\Theta$, $a^2$, $\sigma^2$, $\omega^2$:
\begin{equation}
 T^{(\rm INT)}_{ik} =
-  g_{ik} \nabla_s \left[U^s \left(\Omega_1 \frac{\partial \Phi_*}{\partial \Theta} {+} \Omega_2 \frac{\partial {\cal H}}{\partial \Theta} \right) \right] +
\label{T92}
\end{equation}
$$
{+} 2 DU_i DU_k \left(\Omega_1 \frac{\partial \Phi_*}{\partial a^2} {+} \Omega_2 \frac{\partial {\cal H}}{\partial a^2} \right)  {+}
$$
$$
{+} 2\nabla_s \left\{\left(\Omega_1 \frac{\partial \Phi_*}{\partial a^2} {+} \Omega_2 \frac{\partial {\cal H}}{\partial a^2} \right) \left[DU^s U_iU_k {-} 2U^s U_{(i} DU_{k)} \right] \right\} {-}
$$
$$
- 2\nabla_s\left[\left(\Omega_1 \frac{\partial \Phi_*}{\partial \sigma^2} {+} \Omega_2 \frac{\partial {\cal H}}{\partial \sigma^2} \right) U^s \sigma_{ik} \right] -
$$
$$
{-} 4 \left(\Omega_1 \frac{\partial \Phi_*}{\partial \sigma^2} {+} \Omega_2 \frac{\partial {\cal H}}{\partial \sigma^2} \right)\left[DU_n \sigma^n_{(i} U_{k)} {+} \sigma^n_{(i} \omega_{k)n} \right]{+}
$$
$$
+ 4 \nabla_s \left[\left(\Omega_1 \frac{\partial \Phi_*}{\partial \omega^2} {+} \Omega_2 \frac{\partial {\cal H}}{\partial \omega^2} \right) U_{(i} \omega_{k) \cdot}^{\ \ s} \right] -
$$
$$
{-}4\left(\Omega_1 \frac{\partial \Phi_*}{\partial \omega^2} {+} \Omega_2 \frac{\partial {\cal H}}{\partial \omega^2} \right) \left[DU^n U_{(i} \omega_{k)n} + \sigma^n_{(i} \omega_{k)n}\right]{+}
$$
$$
{+} U_i U_k \left\{ D \left(\Omega_1 \frac{\partial \Phi_*}{\partial \Theta} {+} \Omega_2 \frac{\partial {\cal H}}{\partial \Theta} \right) {-} \right.
$$
$$
 \left. {-} 2 \sigma^2 \left(\Omega_1 \frac{\partial \Phi_*}{\partial \sigma^2}  {+} \Omega_2 \frac{\partial {\cal H}}{\partial \sigma^2} \right){-} \right.
$$
$$
\left. {-}2 \omega^2 \left(\Omega_1 \frac{\partial \Phi_*}{\partial \omega^2}  {+} \Omega_2 \frac{\partial {\cal H}}{\partial \omega^2}\right) {-}
 4 a^2 \left(\Omega_1 \frac{\partial \Phi_*}{\partial a^2} {+} \Omega_2 \frac{\partial {\cal H}}{\partial a^2}\right) \right\} \,.
$$
The fifth term appears, when the system includes some medium; the corresponding stress-energy tensor can be formally written via the variational derivative
\begin{equation}
 T^{(\rm M)}_{ik} = \frac{(-2)}{\sqrt{-g}} \frac{\delta}{\delta g^{ik}} \left[\sqrt{-g} L_{(\rm M)} \right]\,,
\label{T94}
\end{equation}
and can be decomposed algebraically as follows:
\begin{equation}
 T^{(\rm M)}_{ik} = W U_i U_k + I_i U_k + I_k U_i + {\cal P}_{ik} \,.
\label{T929}
\end{equation}
As usual, the scalar $W$ is the energy density of this fluid, $I_k$ is the heat-flux four-vector orthogonal to the aether velocity four-vector, $I_k U^k=0$,  and ${\cal P}_{ik}$ is the symmetric pressure tensor, which also is orthogonal to $U^j$, i.e., $U^i {\cal P}_{ik} {=} 0 {=} {\cal P}_{ik} U^k$.

Let us mention that the main idea to present the given decomposition of the total stress-energy tensor is connected with some convenience; for sure, the terms describing, for instance, the interaction between the aether and electromagnetic field  can not be definitely attributed to the first or second stress-energy tensor.

The Bianchi identity requires that
\begin{equation}
	\nabla^k \left[T^{(\rm U)}_{ik} {+} \kappa T^{(\rm A)}_{ik} {+} \kappa T^{(\rm EM)}_{ik} {+} T^{(\rm INT)}_{ik} {+}  \kappa T^{(\rm M)}_{ik}\right] = 0 \,.
	\label{T99}
\end{equation}
When $T^{(\rm M)}_{ik}=0$, the covariant divergence of the total stress-energy tensor is equal to zero on the solutions of the master equations of the vector, pseudoscalar and electromagnetic fields.
When $T^{(\rm M)}_{ik} \neq 0$, the state functions $W$, $I_k$ and ${\cal P}_{ik}$ are linked by the balance equations
\begin{equation}
	\nabla^k T^{(\rm M)}_{ik} = 0 \,.
	\label{T999}
\end{equation}

\section{The application: Static spherically symmetric axionic dyon}

\subsection{Geometric aspects}

We work with the static spherically symmetric spacetime, which is described by the metric
\begin{equation}
ds^2 = B(r) dt^2 - A(r) dr^2 - r^2 \left(d \theta^2 + \sin^2{\theta} d\varphi^2 \right) \,.
\label{A1}
\end{equation}
The appropriate choice of the aether velocity unit time-like four-vector is
\begin{equation}
U^j = \delta^j_0  \frac{1}{\sqrt{B}} \,, \quad U_j = \delta_j^0 \sqrt{B} \,, \quad  g_{mn}U^m U^n =1 \,.
\label{A2}
\end{equation}
In addition we introduce the unit space-like four-vector orthogonal to the aether velocity four-vector (the director, necessity of which is advocated by G.A. Maugin \cite{MauginJMP})
$$
{\cal R}^k = \delta^k_r  \frac{1}{\sqrt{A}} \,, \quad  {\cal R}_k = - \delta_k^r  \sqrt{A} \,,
$$
\begin{equation}
g_{mn}{\cal R}^m {\cal R}^n =-1 \,, \quad  g_{mn}{\cal R}^m U^n = 0 \,.
\label{A3}
\end{equation}
The covariant derivative of the velocity four-vector takes now the form
\begin{equation}
\nabla_m U_n = - \delta_m^0 \delta_n^r \frac{B^{\prime}}{2\sqrt{B}} \,.
\label{A4}
\end{equation}
Here and below the prime denotes the derivative with respect to radial variable $r$.  The direct calculations show that for this static spherically symmetric field configuration the expansion scalar, the shear and vorticity tensors vanish
\begin{equation}
\Theta = 0 \,, \quad \sigma_{mn} = 0 \,, \quad \omega_{mn} = 0 \,.
\label{A5}
\end{equation}
Only the acceleration four-vector $DU^m$ forms now the covariant derivative
\begin{equation}
\nabla_m U_n =  U_m DU_n \,, \quad DU_n = - \delta_n^r \frac{B^{\prime}}{2B} =  \frac{B^{\prime}}{2B \sqrt{A}} {\cal R}_n \,.
\label{A6}
\end{equation}
Mention should be made that the acceleration four-vector is parallel to the director, and the second order derivative
\begin{equation}
D^2 U_i = U_i \left(\frac{{B^{\prime}}^2}{4AB^2}\right) = - U_i a^2
\label{A7}
\end{equation}
is proportional to the velocity four-vector. Also, one can check directly that
\begin{equation}
\nabla_k DU^k = U^k \nabla_m \nabla_k U^m = U_m R^m_{l} U^l = R^0_0 \,,
\label{A345}
\end{equation}
\begin{equation}
a^2 = - \left(\frac{B^{\prime}}{2 A B} \right)^2
\,,
\label{Ab345}
\end{equation}
\begin{equation}
R^0_0 = \frac{B^{\prime \prime}}{2A B}  + \frac{B^{\prime}}{rA B} - \frac{A^{\prime} B^{\prime}}{4A^2 B} - \frac{{B^{\prime}}^2}{4AB^2} \,.
\label{G111}
\end{equation}
These details will simplify the further calculations.

\subsection{Solution to the key equations for the unit vector field}

The modified tensor (\ref{16}) converts into
$$
{\cal J}^a_{\ j} = [C_1(1+{\cal H})+C_4] U^a DU_j - C_1 U_j DU^a =
$$
\begin{equation}
=\frac{B^{\prime}}{2B\sqrt{A}} \left\{\left[C_1 (1+ {\cal H})+C_4\right]U^a {\cal R}_j - C_1 U_j {\cal R}^a \right\} \,.
\label{A8}
\end{equation}
The reduced version of the equation of the vector field dynamics (\ref{15}) can be presented in the following form:
\begin{equation}
 {-}\frac{1}{A} U_j  C_1 \left[\frac{B^{\prime \prime}}{2B}  {+} \frac{B^{\prime}}{r B} {-} \frac{{B^{\prime}}^2}{4B^2} {-} \frac{A^{\prime} B^{\prime}}{4AB} \right] -
\label{A9}
\end{equation}
$$
- U_j a^2 [C_1(1+{\cal H})+C_4] =
$$
$$
= \lambda  U_j  {+}  \left(C_4 + C_1 {\cal H} \right) U_j a^2 {+}
$$
$$
{+}\kappa {\cal H}\left[U_j E_m E^m - \epsilon_{mjpq}U^q B^p E^m \right] +
$$
$$
+ 4 U_j a^2 \left(\Omega_1 \frac{\partial \Phi_*}{\partial a^2} {+} \Omega_2 \frac{\partial {\cal H}}{\partial a^2}\right) \,.
$$
Below we consider the axionic dyon for which both magnetic and electric fields have only radial components. For such field configuration the term  $\epsilon_{mjpq}U^q B^p E^m$ vanishes, and we see that four equations for the unit vector field reduce to one key equation, which gives us the Lagrange multiplier
$$
 \lambda = - \frac{1}{A} C_1 \left[\frac{B^{\prime \prime}}{2B}  + \frac{B^{\prime}}{r B} - \frac{{B^{\prime}}^2}{4B^2} - \frac{A^{\prime} B^{\prime}}{4AB} \right] -
$$
$$
- a^2 [C_1(1+2{\cal H})+2C_4] -
$$
\begin{equation}
-\kappa {\cal H} E_m E^m - 4  a^2 \left(\Omega_1 \frac{\partial \Phi_*}{\partial a^2} {+} \Omega_2 \frac{\partial {\cal H}}{\partial a^2}\right) \,.
\label{A10}
\end{equation}
Thus, the equations, which describe the vector field configuration, are solved.

\subsection{Solutions to the equations of magneto-electrostatics}

The symmetry of the model under consideration hints us to search for the two-potential solutions to the magneto-electrostatic equations. To be more precise, we assume that
$A_k = \delta_k^0 A_0(r) + \delta_k^{\varphi} A_{\varphi}(\theta)$, and obtain from (\ref{T0}) that
\begin{equation}
A_{\varphi}= Q_{(\rm m)} (1-\cos{\theta}) \,, \quad F_{\theta \varphi} = Q_{(\rm m)} \sin{\theta} \,,
\label{A11}
\end{equation}
where $Q_{(\rm m)}$ is the magnetic charge of the monopole.
Integration of the reduced equations (\ref{T1}) gives only one non-trivial consequence:
\begin{equation}
C^{0rpq} F_{pq} = \frac{K_0}{r^2 \sqrt{AB}} \,,
\label{A12}
\end{equation}
where $K_0$ is a constant of integration. The electric field $E=\sqrt{-E_m E^m}$ can be found from the relation
$$
E = \frac{F_{r0}}{\sqrt{AB}} = \frac{A_0^{\prime}(r)}{\sqrt{AB}} =
$$
\begin{equation}
= \frac{1}{r^2 (1+ {\cal H})} \left[K_0 - \frac{Q_{(\rm m)} \Phi_*}{2\pi} \sin{\left(\frac{2\pi \phi}{\Phi_*}\right)} \right]\,.
\label{A13}
\end{equation}
The first and second invariants of the electromagnetic field are, respectively
$$
\frac14 F_{mn} F^{mn} = \frac12 \left(\frac{Q^2_{(\rm m)}}{r^4}- E^2 \right) \,,
$$
\begin{equation}
\frac14 F^*_{mn} F^{mn} = \frac{1}{r^2 \sqrt{AB}\sin{\theta}}  F_{0r} F_{\theta \varphi} = - \frac{1}{r^2} Q_{(\rm m)} E \,.
\label{A15}
\end{equation}
In the asymptotic limit $r \to \infty$ of the model with the equilibrium axion system $\phi = n \Phi_*$, the formula (\ref{A13}) recovers the Coulomb law $E=\frac{Q_{(\rm e)}}{r^2}$, where $Q_{(\rm e)}$ is the electric charge of the dyon, if $K_0 = Q_{(\rm e)} (1{+} {\cal H}(\infty))$.

\subsection{Reduced equation for the axion field}

The equation for the axion field (\ref{243}) takes the form
$$
\frac{1}{r^2 \sqrt{AB}} \left(r^2 \sqrt{\frac{B}{A}} \phi^{\prime} \right)^{\prime}
 -\frac{m^2_A \Phi_{*}}{2\pi} \sin{\left(\frac{2 \pi \phi}{\Phi_{*}}\right)} =
 $$
 \begin{equation}
 = - \frac{ Q_{(\rm m)} E}{r^2\Psi^2_0}\cos{\left(\frac{2 \pi \phi}{\Phi_{*}}\right)}  \,.
\label{R24}
\end{equation}
Our ansatz is that, when $\phi{=}\Phi_*$, i.e., the axion field is in the second minimum of the periodic potential ($\phi = n \Phi_*$, $n=0,1,2...$), we obtain the key equation for the guiding function of the second type, $\Phi_*$ in the following form:
\begin{equation}
\left(r^2 \sqrt{\frac{B}{A}} \Phi_*^{\prime} \right)^{\prime}
  + \frac{K_0 Q_{(\rm m)}\sqrt{AB}}{r^2 \Psi^2_0 (1+{\cal H})}   =0 \,.
\label{equil1}
\end{equation}

\subsection{Reduced equations for the gravity field}

\subsubsection{The sources of the gravity field}

First of all, we intend to analyze the total stress-energy tensor calculated using the chosen spacetime symmetry.
The term (\ref{326}) reduced for the static spherical symmetry can be written as follows:
$$
	T^{(\rm U)}_{ik} = [C_1(1{+}{\cal H}) {+} C_4]\left[\frac12 g_{ik}  a^2 {+} U_i U_k R^0_0 - DU_i DU_k    \right]
+
$$
\begin{equation}
+ C_1 U_i U_k DU^m \nabla_m {\cal H} \,,
\label{G1}
\end{equation}
or equivalently
$$
T^{(\rm U) i}_{\ \ \ k} = [C_1(1+{\cal H})+C_4] \times
$$
\begin{equation}
\times \left\{
 - \left(\frac12 \delta^i_{k} + {\cal R}^i {\cal R}_k + U^iU_k \right) \frac{{B^{\prime}}^2}{4AB^2}    +  \right.
 \label{26}
\end{equation}
$$
\left. + U^iU_k \left[\frac{B^{\prime \prime}}{2A B}  + \frac{B^{\prime}}{rA B} - \frac{A^{\prime} B^{\prime}}{4A^2 B} \right] \right\} -
$$
$$
-  C_1 U^i U_k \frac{\partial {\cal H}}{\partial a^2} \left(\frac{B^{\prime}}{2AB}\right)  \left[\frac{{B^{\prime}}^2}{4AB^2}\right]^{\prime} \,.
$$
The terms (\ref{T76}), (\ref{T91}) can be rewritten, respectively, as
\begin{equation}
	T^{(\rm A)i}_{\ \ \ \ k} = \Psi^2_0 \left[\left( \frac12 \delta^i_k - \delta^i_r \delta_k^r\right) \frac{(\phi^{\prime})^2}{A}  + \frac12 \delta^i_{k} V   \right]\,,
		\label{TA3}
\end{equation}
$$
 T^{(\rm EM) i}_{\ \ \ \ \ k}=  (1 + {\cal H}) E^2 \left(-\frac12 \delta^i_{k} + \delta^i_0 \delta_k^0 + \delta^i_r \delta_k^r \right) +
$$
\begin{equation}
+  \frac{Q^2_{(\rm m)}}{r^4}\left(\frac12 \delta^i_{k} - \delta^i_{\theta} \delta_k^{\theta} - \delta^i_{\varphi} \delta_k^{\varphi} \right)\,.
\label{TA4}
\end{equation}
The interaction term (\ref{T92}) is simplified essentially:
\begin{equation}
 T^{i(\rm INT)}_{k} = - 4 a^2 U^iU_k   \left(\Omega_1 \frac{\partial \Phi_*}{\partial a^2} {+} \Omega_2 \frac{\partial {\cal H}}{\partial a^2}\right)  +
\label{TInt6}
\end{equation}
$$
+ 2 DU^i DU_k \left(\Omega_1 \frac{\partial \Phi_*}{\partial a^2} {+} \Omega_2 \frac{\partial {\cal H}}{\partial a^2} \right)  +
$$
$$
+ 2\nabla_s \left\{\left(\Omega_1 \frac{\partial \Phi_*}{\partial a^2} {+} \Omega_2 \frac{\partial {\cal H}}{\partial a^2} \right) \times \right.
$$
$$
\left. \times \left[DU^s U^iU_k - U^s U^{i} DU_{k} - U^s U_{k} DU^{i} \right] \right\} =
$$
$$
= 2 \left(\Omega_1 \frac{\partial \Phi_*}{\partial a^2} {+} \Omega_2 \frac{\partial {\cal H}}{\partial a^2} \right)\left(\delta^i_0 \delta_k^0 R^0_0 - \delta^i_r \delta_k^r a^2  \right) {+}
$$
$$
+ 2\delta^i_0 \delta_k^0 DU^r \frac{d}{dr}\left(\Omega_1 \frac{\partial \Phi_*}{\partial a^2} {+} \Omega_2 \frac{\partial {\cal H}}{\partial a^2} \right) =
$$
$$
= 2 \left(\Omega_1 \frac{\partial \Phi_*}{\partial a^2} {+} \Omega_2 \frac{\partial {\cal H}}{\partial a^2} \right) \times
$$
$$
\times \left[\delta^i_0 \delta_k^0 \left(\frac{B^{\prime \prime}}{2A B}  {+} \frac{B^{\prime}}{rA B} {-} \frac{A^{\prime} B^{\prime}}{4A^2 B} {-} \frac{{B^{\prime}}^2}{4AB^2}\right) {+} \delta^i_r \delta_k^r \frac{{B^{\prime}}^2}{4AB^2} \right] {+}
$$
$$
{+} \delta^i_0 \delta_k^0 \frac{B^{\prime}}{AB} \frac{d}{dr}\left(\Omega_1 \frac{\partial \Phi_*}{\partial a^2} {+} \Omega_2 \frac{\partial {\cal H}}{\partial a^2} \right) \,.
$$

\subsubsection{Key equations for the gravity field}

For the static spherically symmetric configurations with vanishing cosmological constant, $\Lambda{=}0$, there are two independent gravity field equations; we prefer to work with the Einstein equations related to the terms $G^0_0$ and $G^r_r$, respectively, in the left-hand sides. Thus, the first key equation is
$$
\frac{1}{r^2 A}(A-1) + \frac{A^{\prime}}{r A^2} =
$$
\begin{equation}
= [C_1(1{+}{\cal H}){+}C_4]\left({-}\frac{3{B^{\prime}}^2}{8AB^2} {+} \frac{B^{\prime \prime}}{2A B} {+} \frac{B^{\prime}}{rA B} {-} \frac{A^{\prime} B^{\prime}}{4A^2 B} \right) {-}
\label{KEY1}
\end{equation}
$$
- C_1 \left(\frac{B^{\prime}}{2A B} \right) \left(\frac{{B^{\prime}}^2}{4AB^2}\right)^{\prime} \frac{\partial {\cal H}}{\partial a^2} +
$$
$$
+ \frac12 \kappa \Psi^2_0 \left( V + \frac{1}{A} {\phi^{\prime}}^2  \right) + \frac{\kappa}{2} \left[\frac{Q^2_{(\rm m)}}{r^4} + (1+{\cal H})E^2 \right] +
$$
$$
+2 \left(\Omega_1 \frac{\partial \Phi_*}{\partial a^2} {+} \Omega_2 \frac{\partial {\cal H}}{\partial a^2} \right)\left[ \left(\frac{B^{\prime \prime}}{2A B}  {+} \frac{B^{\prime}}{rA B} {-} \frac{A^{\prime} B^{\prime}}{4A^2 B} {-} \frac{{B^{\prime}}^2}{4AB^2}\right) \right] +
$$
$$
{+} \frac{B^{\prime}}{AB} \frac{d}{dr}\left(\Omega_1 \frac{\partial \Phi_*}{\partial a^2} {+} \Omega_2 \frac{\partial {\cal H}}{\partial a^2} \right) \,,
$$
and the second key equation for the gravity field takes the form
\begin{equation}
\frac{1}{r^2 A}(A-1) - \frac{B^{\prime}}{r AB} = [C_1(1+{\cal H})+C_4]\frac{{B^{\prime}}^2}{8AB^2} +
\label{KEY2}
\end{equation}
$$
+\frac12 \kappa \Psi^2_0 \left( V - \frac{1}{A} {\phi^{\prime}}^2  \right) + \frac{\kappa}{2} \left[\frac{Q^2_{(\rm m)}}{r^4} + (1+{\cal H})E^2 \right] +
$$
$$
+ 2 \left(\Omega_1 \frac{\partial \Phi_*}{\partial a^2} {+} \Omega_2 \frac{\partial {\cal H}}{\partial a^2} \right)\left( \frac{{B^{\prime}}^2}{4AB^2} \right) \,.
$$

\subsubsection{Short resume}

Working with the static spherically symmetric model we have to solve four key equations (\ref{24}),  (\ref{equil1}), (\ref{KEY1}) and (\ref{KEY2}) for four unknown functions $\phi(r)$, $\Phi_*(r)$, $B(r)$ and $A(r)$ with $E(r)$ given by (\ref{A13}). As for the guiding function ${\cal H}(a^2)$, it remains to be modeled. The study of the mentioned system of equations requires the use of qualitative and numerical analysis similar to the work done by the authors of the papers \cite{AS1,AS2} for the models without axions and magnetic field. This work is beyond the scope of this article, but we hope to do it in the nearest future.

\subsection{Limiting case: The solution of the Reissner-Nordstr\"om type  }

The developed extended model should have the special  case, describing the Reissner-Nordstr\"om solution, for which $A(r) \cdot B(r)=1$. When $A=\frac{1}{B}$ the difference of the left-hand sides of the equations (\ref{KEY1}) and (\ref{KEY2}) vanishes, thus the functions ${\cal H}(a^2)$ and $\Phi_*(a^2)$ should satisfy the condition that the difference of the right-hand sides also be vanishing. This is possible, in particular, when $C_1 \neq 0$ and the function ${\cal H}$ is equal to the constant ${\cal H} = - \frac{C_1{+}C_4}{C_1}$. Also, we have to require that the axion field is frozen in the lowest level of the potential $V$ (see (\ref{13})), i.e., $\phi {=} 0$. In its turn, according to (\ref{24}), it is possible, when $E=0$, i.e., $K_0=0$ according to (\ref{A13}). Since the case $\phi {=}0$ relates to the equilibrium state of the axion system, we see that $V(0)=0$ and $\Omega_1(0)=0$. As the result of these requirements, we obtain the equation for $A(r)$
\begin{equation}
\frac{1}{r^2 A}(A-1) + \frac{A^{\prime}}{r A^2} = \frac{\kappa Q^2_{(\rm m)}}{2r^4}\,,
\label{KEYrn1}
\end{equation}
the solution to which is known as the Reissner-Nordstr\"om solution
\begin{equation}
B(r) = \frac{1}{A(r)} = 1- \frac{r_g}{r} + \frac{r^2_{Q}}{r^2} \,,
\label{KEYrn2}
\end{equation}
where $r_{g} {=} \frac{2GM}{c^2}$ is the Schwarzschild radius, and $r^2_{Q}{=} \frac{4\pi G Q^2_{(\rm m)}}{c^4}$ is the square of the Reissner-Nordstr\"om radius, $M$ is the asymptotic mass of the object, $G$ is the Newtonian gravitational constant, $c$ is the speed of light in vacuum. The square of the acceleration four-vector can be calculated using the metric (\ref{KEYrn2})
\begin{equation}
a^2 {=} {-} \frac{{B^{\prime}}^2}{4B} = {-} \left(\frac{r^2_{g}}{4r^4}\right) \frac{\left(r{-}\frac{2 r^2_{Q}}{r_{g}} \right)^2}{\left[\left(r{-}\frac12 r_{g} \right)^2 {+} \left(r^2_{Q}{-}\frac{1}{4}r^2_{g}\right)\right]} \,.
\label{eqRN71}
\end{equation}
The guiding function of the second type $\Phi_*$ is now the solution to the equation
\begin{equation}
\left(r^2 B \Phi_*^{\prime} \right)^{\prime}{=}0  \Rightarrow \Phi_*^{\prime}(r) {=} \frac{\alpha}{\left[\left(r{-}\frac12 r_{g} \right)^2 {+} \left(r^2_{Q}{-}\frac14 r^2_{g} \right)\right]} .
\label{eqRN7}
\end{equation}
The constant integration $\alpha$ can be reformulated as
\begin{equation}
\alpha = \lim_{r \to \infty}[r^2 \Phi_*^{\prime}(r)] \,.
\label{eqRN9}
\end{equation}
Depending on the ratio of the radii $r_{g}$ and $r_{Q}$, there are three versions of the representation of the function $\Phi_{*}$.

1. When $r_{Q}>\frac12 r_{g}$, the metric (\ref{KEYrn2}) is regular. The square of the acceleration four-vector (\ref{eqRN71}) is regular and negative for $0<r< \infty$. When $r \to \infty$, the function $-a^2$ asymptotically vanishes $\propto \frac{1}{r^4}$; when $r = \frac{2 r^2_{Q}}{r_{g}}$, this function takes zero value. Clearly, the graph of this function has the minimum at $r=r_{(\rm min)}$. The corresponding solution for the guiding function of the second type is also regular
\begin{equation}
\Phi_*(r) {=} \Phi_*(\infty) {+} \frac{\alpha}{\sqrt{r^2_{Q}{-}\frac14 r^2_{g}}}\left\{\arctan{\left[\frac{\left(r{-}\frac12 r_{g} \right)}{\sqrt{r^2_{Q}{-}\frac14 r^2_{g}}}\right]} {-} \frac{\pi}{2}\right\}.
\label{eqRN8}
\end{equation}
The pair of the functions  $\Phi_*(r)$ (\ref{eqRN8}) and $a^2(r)$ (\ref{eqRN71}) gives the parametric representation of the guiding function of the second type $\Phi_*(a^2)$. There are no chances to find   $\Phi_*(a^2)$ analytically, but it is easy to reconstruct the corresponding profile numerically.

2. When $r_{Q}< \frac12 r_{g}$, the metric function $B(r)$ can be written in well-known form
\begin{equation}
B(r) = \left(1{-} \frac{r_{+}}{r} \right)\left(1{-} \frac{r_{-}}{r} \right) \,, \quad r_{\pm} = \frac12 r_{g} \pm \sqrt{\frac14 r^2_{g}{-} r^2_{Q}} \,,
\label{hor1}
\end{equation}
where the parameters $r_{+}$ and $r_{-}$ are the radii of the external and internal horizons, respectively.
Now the guiding function of the second type is of the logarithmic form:
\begin{equation}
\Phi_*(r) = \Phi_*(\infty) +\frac{\alpha}{(r_{+}-r_{-})} \log{\left|\frac{r-r_{+}}{r-r_{-}} \right|} \,.
\label{eqRN802}
\end{equation}
Clearly, the functions $A(r)$, $a^2(r)$ and $\Phi_*(r)$ become infinite on the horizons, and are  regular in the zone $r>r_{+}$.

3. When $r_{Q}{=} \frac12 r_{g}$, we deal with the case $r_{+}{=} r_{-}{=}r_{Q}$, and the metric coefficient $B(r)= \left(1{-}\frac{r_{Q}}{r}\right)^2$ is non-negative and demonstrates the presence of one (double) horizon. Now the guiding function is of the simple form
\begin{equation}
\Phi_*(r) = \Phi_*(\infty) +\frac{\alpha}{(r_{Q}-r)} \,.
\label{eqRN80}
\end{equation}
Since now $a^2 = - \left(\frac{r_{Q}}{r^2}\right)^2$, we can reconstruct the function $\Phi_*(a^2)$ analytically:
\begin{equation}
\Phi_*(a^2) = \Phi_*(0) + \frac{\frac{\alpha}{r_{Q}}}{\left[1- (-a^2 r^2_{Q})^{-\frac14} \right]} \,.
\label{eqRN806}
\end{equation}
Again, the guiding function tends to infinity, when the observer approaches to the double horizon $r=r_{Q}$, or equivalently $a^2 r^2_{Q} = {-} 1$, and tends to zero on the spatial infinity, when $r \to \infty$ and $a^2 \to 0$.

Finally, the classical Reissner-Nordstr\"om formula relates to the particular solution to the set of master equations for the extended Einstein-Maxwell-aether-axion model, when  the guiding function of the first type is constant and can be expressed in terms of the Jacobson's coupling constant as  ${\cal H} = {-} \left(\frac{C_1{+}C_4}{C_1} \right)$, and the guiding function of the second type, $\Phi_*$, is presented by one of the formulas (\ref{eqRN8}), (\ref{eqRN802}), (\ref{eqRN80}), depending on the value of the ratio $\frac{r_{g}}{r_{Q}}$.

\section{Discussion and conclusions}

We have elaborated the covariant formalism, which allows us to describe a two-level aetheric control over the evolution (distribution) of the axionically active electrodynamic systems.
We indicate this theory as the extended Einstein-Maxwell-aether-axion theory. This extension is two folds. The first idea was to modify the kinetic terms for the vector, pseudoscalar and electromagnetic fields by replacing the physical metric $g^{ik}$ with the aetheric effective metric $G^{ik}=g^{ik}{+} {\cal H} U^i U^k$ in the corresponding constitutive tensors. We indicated the appeared new function ${\cal H}$ as the guiding function of the first type; it depends on four scalars, $\Theta$, $a^2$, $\sigma^2$, $\omega^2$, constructed using the covariant derivative of the aether velocity four-vector (see (\ref{F54})-(\ref{06})). The second idea concerned the modifications of the potential of the axion field $V(\phi, \Phi_*)$ and of the contact term, describing the axion-photon coupling. On this way, the guiding function of the second type, $\Phi_*(\Theta,a^2,\sigma^2,\omega^2)$ appeared. As the result, the modified action functional has been proposed in the form (\ref{00}).

Clearly, the arguments of both guiding functions depend on the aether velocity four-vector $U^j$ and on the metric $g^{ik}$. This means that the procedure of variation with respect to $U^j$ and  $g^{ik}$ extends all the master equations of the model (see Section IIB1 for the aether velocity, Section IIB2 for the axion field, Section IIB3 for the electromagnetic field and IIB4 for the gravity field).

Of course the reader may have a question: why all these complex mathematical calculations are needed? This complex approach is predetermined by various applications to cosmology and astrophysics.
For instance, for the static spherically symmetric models $\Theta{=}0$, $\sigma_{mn}{=}0$, $\omega_{mn}{=}0$, and we can operate with guiding functions depending on the argument $a^2$ only, i.e., ${\cal H}(a^2)$ and $\Phi_{*}(a^2)$. When we deal with the isotropic FLRW type cosmology, only the guiding functions ${\cal H}(\Theta)$ and $\Phi_{*}(\Theta)$ are admissible. The cosmological models of the G\"odel type should operate with ${\cal H}(\omega^2)$ and $\Phi_{*}(\omega^2)$. The anisotropic cosmological models of the Bianchi type, as well as, the models with gravitational waves, require the guiding function to be of the form ${\cal H}(\Theta,\sigma^2)$ and $\Phi_{*}(\Theta,\sigma^2)$. In other words, we are working on the wide program, and this complex approach seems to be reasonable.

In the second part of this paper we applied the extended model for description of the static spherically symmetric dyon in order to show how does this approach work. There is a number of interesting sub-models in this context, however, we hope to consider them in the next papers.

\acknowledgments{The work was supported by the Russian Science Foundation (Grant No 21-12-00130).}


\section*{References}

\begin{thebibliography}{99}

\bibitem{Gordon} W. Gordon, {\it Ann. Phys. (Leipzig)} {\bf 72}, 421, (1923).

\bibitem{GR} J.L. Singe, "Relativity: The general theory", North-Holland P.C., Amsterdam, 1960.

\bibitem{PMQ} Pham Mau Quan, {\it Arch. Rat. Mech. Anal.} {\bf 1},  54 (1957).

\bibitem{MauginJMP} G.A. Maugin, {\it J. Math. Phys.} {\bf 19}, 1206 (1978).

\bibitem{HehlObukhov} F.W. Hehl, Yu.N. Obukhov, "Foundations of Classical
Electrodynamics: Charge, Flux, and Metric", Birkh\"auser, Boston,
2003.

\bibitem{LL} L.D. Landau, E.M. Lifchitz, L.P. Pitaevskii,
"Electrodynamics of Continuous Media", Butterworth Heinemann, Oxford,
1996.

\bibitem{BZ} A.B. Balakin and W. Zimdahl, {\it Gen. Relativ. Gravit.} {\bf 37},  1731 (2005).

\bibitem{Visser1} C. Barcel\'o, S. Liberati, M. Visser, {\it Living Rev. Rel.} {\bf 8}, 12 (2005).

\bibitem{Novello} M. Novello, M. Visser, G. Volovik (Eds.),
"Artificial Black Holes", World Scientific, Singapore, 2002.

\bibitem{VolovikBook} G.E. Volovik, "The Universe in a Helium
Droplet", Clarendon, Oxford, 2003.

\bibitem{Volovik} G.E. Volovik, {\it Phys. Rept.} {\bf 351},  195 (2001).

\bibitem{BDZ1} A.B. Balakin A.B., H. Dehnen and A.E. Zayats, {\it Phys. Rev. D} {\bf 76}, 124011 (2007).

\bibitem{BDZ2} A.B. Balakin A.B., H. Dehnen and A.E. Zayats, {\it Ann. Phys.} {\bf 323} 2183 (2008).

\bibitem{BDZ3} A.B. Balakin A.B., H. Dehnen and A.E. Zayats, {\it Gen. Relat. Grav.} {\bf 40}, 2493 (2008).

\bibitem{J1} T. Jacobson and D. Mattingly, {\it Phys. Rev. D} {\bf 64} 024028 (2001).

\bibitem{J2}  T. Jacobson, {\it PoSQG-Ph} {\bf 020},  020 (2007).

\bibitem{J3} T. Jacobson and D. Mattingly, {\it Phys. Rev. D} {\bf 70},  024003 (2004).

\bibitem{J4} C. Heinicke, P. Baekler and F.W. Hehl,  {\it Phys. Rev. D} {\bf 72} 025012 (2005).

\bibitem{J5} A.B. Balakin and J.P.S. Lemos,  {\it Ann. Phys.} {\bf 350},  454 (2014).

\bibitem{PQ} R.D. Peccei and H.R. Quinn,  {\it Phys. Rev. Lett.} {\bf 38}, 1440-1443 (1977).

\bibitem{Weinberg} S. Weinberg, {\it Phys. Rev. Lett.} {\bf 40}, 223-226 (1978).

\bibitem{Wilczek} F. Wilczek,  {\it Phys. Rev. Lett.} {\bf 40}, 279 (1978).

\bibitem{WTNi77} Wei-Tou Ni,   {\it Phys. Rev. Lett.} {\bf 38},  301 (1977).

\bibitem{a1}  P. Sikivie, {\it Phys. Rev. Lett.} {\bf 51}, 1415 (1983).

\bibitem{a2} F. Wilczek, {\it Phys. Rev. Lett.} {\bf 58}, 1799 (1987).

\bibitem{ADM1}  L.D. Duffy and  K. van Bibber,  {\it New J. Phys.} {\bf 11}, 105008 (2009).

\bibitem{ADM2} M. Khlopov, "Fundamentals of Cosmic Particle Physics", CISP-Springer: Cambridge, UK, 2012.

\bibitem{ADM3} A. Del Popolo, {\it Int. J. Mod. Phys. D} {\bf 23}, 1430005 (2014).

\bibitem{DM} G. Bertone, D. Hooper and J. Silk, {\it Phys. Rept.}, {\bf 405},  279 (2005).

\bibitem{a5}
D.J.E. Marsh,  {\it Phys. Rept.} {\bf 643}, 1 (2016).

\bibitem{GW}
LIGO Scientific Collaboration, Virgo Collaboration, Fermi Gamma-Ray Burst Monitor, INTEGRAL. {\it Astrophys. J. Lett.} {\bf 848}, L13 (2017).

\bibitem{Equilib1} A.B. Balakin and A.F. Shakirzyanov,  {\it Phys. Dark Univ.}, {\bf 24}, 100283 (2019).

\bibitem{Equilib2} A.B. Balakin and D.E. Groshev,  {\it Eur. Phys. J. C} {\bf 80},  145 (2020).

\bibitem{Equilib3} A.B. Balakin and D.E. Groshev, {\it Symmetry}  {\bf 12}(3), 455  (2020).

\bibitem{Equilib4} A.B. Balakin and D.E. Groshev, {\it Int. J. Mod. Phys. D} {\bf 29}, 2050083 (2020).

\bibitem{Equilib5} A.B. Balakin and A.F. Shakirzyanov,  {\it Universe} {\bf 6}(11), 192 (2020).

\bibitem{oiko1} V.K. Oikonomou,  {\it Phys. Rev. D} {\bf 106}, 044041 (2022).

\bibitem{oiko2}  S.D. Odintsov, and V.K. Oikonomou,  {\it Phys. Rev. D} {\bf 99}, 104070 (2019).

\bibitem{AS1} C. Eling and T. Jacobson,{\it Class. Quant. Grav.} {\bf 23}, 5643, (2006).

\bibitem{AS2} C. Eling, T. Jacobson and M.C. Miller, {\it Phys. Rev. D} {\bf 76},  042003, (2007).

\end{thebibliography}
\end{document}